\documentclass[pra,aps,twocolumn,superscriptaddress,showpacs,floatfix]{revtex4}
\usepackage{amsmath}
\usepackage{graphicx}

\def\prn#1{{\left(#1\right)}}
\def\cbrk#1{{\left\{#1\right\}}}
\def\sbrk#1{{\left[#1\right]}}

\def\rightbar#1{\left.#1\right|}
\def\bra#1{{\left\langle#1\right\vert}}
\def\ket#1{{\left\vert#1\right\rangle}}
\def\cg#1#2{{\left\langle#1\vert#2\right\rangle}}
\def\threej(#1,#2)(#3,#4)(#5,#6){\begin{pmatrix}#1&#3&#5\\#2&#4&#6\end{pmatrix}}
\def\sixj(#1,#2,#3)(#4,#5,#6){\begin{Bmatrix}#1&#2&#3\\#4&#5&#6\end{Bmatrix}}
\def\ninej(#1,#2,#3)(#4,#5,#6)(#7,#8,#9){\begin{Bmatrix}#1&#2&#3\\#4&#5&#6\\#7&#8&#9\end{Bmatrix}}

\begin{document}

\setkeys{Gin}{width=3.25 in}

\title{Nonlinear magneto-optical rotation in optically thick media}
\author{S. M. Rochester}
\affiliation{Department of Physics, University of California at
Berkeley, Berkeley, California 94720-7300}
\author{D. Budker}
\email{budker@socrates.berkeley.edu} \affiliation{Department of
Physics, University of California at Berkeley, Berkeley,
California 94720-7300} \affiliation{Nuclear Science Division,
Lawrence Berkeley National Laboratory, Berkeley, California 94720}
\date{\today}

\begin{abstract}
Nonlinear magneto-optical rotation is a sensitive technique for
measuring magnetic fields. Here, the shot-noise-limited
magnetometric sensitivity is analyzed for the case of
optically-thick media and high light power, which has been the
subject of recent experimental and theoretical investigations.
\end{abstract}



\maketitle

\section{Introduction}

Resonant nonlinear magneto-optical rotation (NMOR)
\cite{Gaw94,Bud99} has been the subject of extensive theoretical
and experimental studies because it provides a very sensitive way
to measure magnetic fields (see, e.g., Ref.\ \cite{Bud2000Sens}).
While NMOR experiments are usually carried out with vapor samples
of moderate optical thickness (no more than $\sim 2$ absorption
lengths), the intense research of recent years on
electromagnetically-induced transparency (EIT; see, e.g., Ref.\
\cite{Har97} for a review) has motivated investigation of
optically thick media \cite{Scu92,Fle94}.

Recently, NMOR in the vicinity of the D1 and D2 lines by an
optically thick vapor of rubidium was studied \cite{Sau2000,
NovLargePol2001,MatRadTrap2001,Nov2002Mag,Mat2002Rad}. The authors
of Ref.\ \cite{NovLargePol2001} used a $5\ $cm-long vapor cell
containing $^{87}$Rb and a laser beam tuned to the maximum of the
NMOR spectrum near the D1 line (laser power was $2.5$ mW, beam
diameter $\sim2$ mm). They measured maximum (with respect to laser
frequency and magnetic field) polarization rotation
$\varphi_{max}$ as a function of atomic density. It was found that
$\varphi_{max}$ increases essentially linearly up to $n\approx
3.5\times 10^{12}{\rm\ cm^{-3}}$. At higher densities,
$d\varphi_{max}/dn$ decreases and eventually becomes negative. The
maximum observed rotation was $\approx 10\ $rad with the applied
magnetic field of $\approx 0.6\ $G. (This shows the effect of
power broadening on the magnetic field dependence; at low light
power, the maximum rotation would occur at a magnetic field about
an order of magnitude smaller.) Rotation slope
$d\varphi/dB|_{B=0}$, an important parameter for magnetometry, was
also measured as a function of atomic density in Ref.\
\cite{NovLargePol2001}.

The goal of the present contribution is to analyze the scaling of
optimal magnetometric sensitivity with respect to optical
thickness of the sample. We analyze NMOR for a simple
system\textemdash an isolated $J=1\rightarrow J=0$
transition\textemdash for which analytical solutions for the
density matrix are readily obtained. We assume that the transverse
thickness of the sample is small, so that the trapping of
spontaneously-emitted radiation can be neglected. (This assumption
may not be justified in actual experiments with optically thick
media \cite{MatRadTrap2001}.) We also neglect mixing between
different velocity groups due to velocity-changing collisions.
Such mixing is important in the presence of a buffer gas, or for
anti-relaxation-coated cells.

A theoretical treatment of NMOR in dense media has been carried
out earlier in Refs.\ \cite{Sau2000} and \cite{Fle2000}. The
present analysis partially overlaps with this earlier work and
extends it in several ways: we extend the treatment to the case of
large Doppler broadening and the intermediate regime where power
and Doppler widths are comparable; provide a qualitative
discussion of the contribution to optical rotation as a function
of optical depth in the medium; discuss the role of the coherence
and the alignment-to-orientation conversion effects
\cite{Bud2000AOC}; discuss scaling of magnetometric sensitivity
with optical thickness of the medium; and finally, give a general
argument on scaling based on the observation that an optimized
nonlinear optical rotation measurement is a way to carry out a
spin measurement of an atomic ensemble with sensitivity given by
fundamental quantum noise limits \cite{Bud2000Sens,Fle2000}.

\section{Description of the density matrix calculation}

The calculation is based on a standard density matrix approach.
The time evolution of the atomic density matrix $\rho$ is given by
the Liouville equation (see, e.g., Ref.\ \cite{Ste84}):
\begin{align}
  \frac{d\rho}{dt}=\frac{1}{i\hbar}\sbrk{H,\rho}-\frac{1}{2}\cbrk{\Gamma,\rho}+\Lambda,
  \label{Louiville}
\end{align}
where the square brackets denote the commutator and the curly
brackets the anti-commutator, and the total Hamiltonian $H$ is the
sum of light-atom interaction Hamiltonian
$H_L=-\vec{d}\cdot\vec{E}$ (where $\vec{E}$ is the electric field
vector, and $\vec{d}$ is the electric dipole operator), the
magnetic field-atom interaction Hamiltonian
$H_B=-\vec{\mu}\cdot\vec{B}$ (where $\vec{B}$ is the magnetic
field and $\vec{\mu}$ is the magnetic moment), and the unperturbed
Hamiltonian $H_0$. $\Gamma$ is the relaxation matrix (diagonal in
the collision-free approximation)
\begin{align}
  \bra{\xi Jm}\Gamma\ket{\xi Jm}=\gamma+\gamma_0\delta\prn{\xi,\xi_e},
\end{align}
where $\gamma$ is the ground state depolarization rate (e.g., due
to transit of atoms through the laser beam), $\gamma_0$ is the
spontaneous decay rate from the upper state, and $\xi$ represents
the quantum number distinguishing the ground state ($\xi_g$) from
the excited state ($\xi_e$). $\Lambda=\Lambda^0+\Lambda^{repop}$
is the pumping term, where the diagonal matrix
\begin{align}
   \bra{\xi_gJ_gm}\Lambda^0\ket{\xi_gJ_gm}=\frac{\gamma\rho_0}{\prn{2J_g+1}}
\end{align}
describes incoherent ground state pumping ($\rho_0$ is the atomic
density), and
\begin{widetext}
\begin{align}
    \bra{\xi_gJ_gm}\Lambda^{repop}\ket{\xi_gJ_gm'}=\gamma_0\sum_{m_e,m_e',q}\cg{J_g,m,1,q}{J_e,m_e}\cg{J_g,m',1,q}{J_e,m_e'}\rho_{\xi_eJ_em_e\xi_eJ_em_e'},
\end{align}
\end{widetext}
describes repopulation due to spontaneous relaxation from the
upper level (see, e.g., Ref.\ \cite{Rautian}). Here
$\cg{\ldots}{\ldots}$ are the Clebsch-Gordan coefficients.

The electric field vector is written (see, e.g., Ref.\
\cite{Huard})
\begin{align}
  \vec{E}&=\frac{1}{2}\sbrk{E_0e^{i\phi}\prn{\cos\varphi\cos\epsilon-i\sin\varphi\sin\epsilon}e^{i\prn{\omega
  t-kz}}+c.c.}\hat{x}\nonumber\\
  &+\frac{1}{2}\sbrk{E_0e^{i\phi}\prn{\sin\varphi\cos\epsilon+i\cos\varphi\sin\epsilon}e^{i\prn{\omega
t-kz}}+c.c.}\hat{y},\nonumber\\ \label{lightfield}
\end{align}
where $\omega$ is the light frequency, $k=\omega/c$ is the vacuum
wave number, $E_0$ is the electric field amplitude, $\varphi$ is
the polarization angle, $\epsilon$ is the ellipticity (arctangent
of the ratio of the major and minor axes of the polarization
ellipse), and $\phi$ is the overall phase. By substituting
(\ref{lightfield}) into the wave equation
\begin{align}
  \prn{\frac{\omega^2}{c^2}+\frac{d^2}{dz^2}}\vec{E}=-\frac{4\pi}{c^2}\frac{d^2}{dt^2}\vec{P},
\end{align}
where $\vec{P}=Tr(\rho\vec{d})$ is the polarization of the medium,
the absorption, rotation, phase shift, and change of ellipticity
per unit distance for an optically thin medium can be found in
terms of the density matrix elements (these expressions are given
in Ref.\ \cite{Roc2001SR}). Once the solutions for the density
matrix are obtained, we will perform an integration to generalize
the result to media of arbitrary thickness.

\section{The Doppler-free case}

We consider the case of a $J=1\rightarrow J=0$ transition, and
linearly-polarized incident light, with a magnetic field directed
along the light propagation direction (Faraday geometry). Using
the rotating wave approximation, the solution of Eq.\
(\ref{Louiville}) is obtained, and from this, analytic expressions
for thin medium absorption and rotation are found. These
expressions can be simplified by assuming that
$\gamma\ll\gamma_0$. We first consider the case where the
power-broadened line width is much greater than the Doppler width.
In this case, the absorption coefficient per unit length $\alpha$
is found to be
\begin{align}
    \alpha\approx\frac{\alpha_0}{(2\Delta/\gamma_0)^2+2\kappa/3+1},
    \label{AlphaDF}
\end{align}
where $\Delta$ is the light-frequency detuning from resonance,
\begin{align}
\kappa = \frac{d^2E_0^2}{\hbar^2\gamma\gamma_0}
\end{align}
is the optical pumping saturation parameter, and
\begin{align}
    \alpha_0\approx\frac{1}{6\pi}\lambda^2 n
\end{align}
is the unsaturated absorption coefficient on resonance, where
$\lambda$ is the transition wavelength and $n$ is the atomic
density. For $\Omega\ll\gamma$, where $\Omega=g \mu B$ is the
Larmor frequency ($g$ is the Land\'{e} factor, and $\mu$ is the
Bohr magneton), the slope of optical rotation per unit length,
$d\varphi/\prn{d\Omega dx}$, (proportional to rotation for small
magnetic fields) is found to be
\begin{align}
    \frac{d\varphi}{d\Omega
    dx}\approx\frac{1}{\gamma}\frac{\alpha_0}{(2\Delta/\gamma_0)^2+2\kappa/3},
\label{Slopedx}
\end{align}
where we have neglected linear optical rotation. In general,
optical rotation can be induced by either linear dichroism or
circular birefringence. Analysis of the steady-state polarization
of the ground state shows that both processes contribute here: the
contribution due to linear dichroism induced in the medium is
given by
\begin{align}
   \rightbar{\frac{d\varphi}{d\Omega dx}}_{dichr.}\approx \frac{1}{\gamma}
   \frac{1}{1+\prn{2\Delta/\gamma_0}^2}
   \frac{\alpha_0}{(2\Delta/\gamma_0)^2+2\kappa/3},
\end{align}
and the contribution due to circular birefringence (arising due to
alignment-to-orientation conversion in the presence of both the
magnetic field and the strong electric field of the light
\cite{Bud2000AOC}) is given by
\begin{align}
   \rightbar{\frac{d\varphi}{d\Omega dx}}_{biref.}\approx \frac{1}{\gamma}
    \frac{\prn{2\Delta/\gamma_0}^2}{1+\prn{2\Delta/\gamma_0}^2}
    \frac{\alpha_0}{(2\Delta/\gamma_0)^2+2\kappa/3}.
\end{align}
The sum of the two contributions produces the Lorentzian line
shape of Eq.\ (\ref{Slopedx}).

We now generalize the formulas for absorption and rotation to the
case of thick media, and find the magnetometric sensitivity. In
the Doppler-free case we are presently considering, we can further
simplify the expressions by assuming $\Delta=0$ (we will have to
include non-zero detunings in the discussion of the
Doppler-broadened case below), and $\kappa\gg1$. (We will see from
the final result that this holds everywhere in the medium when the
input light power is optimal.) Generally, the medium in the
presence of a magnetic field produces light ellipticity as well as
rotation; however, the ellipticity is an odd function of detuning
and is zero on resonance. The rate of change of the saturation
parameter as light travels through the medium is given by
\begin{align}
    \frac{d\kappa(x)}{dx}&=-\alpha\kappa(x)\nonumber\\
    &\approx-\frac{3}{2}\alpha_0,
\end{align}
so solving for the saturation parameter as a function of position,
\begin{align}
    \kappa(x)\approx\kappa_0-\frac{3}{2}\alpha_0x,
    \label{kappaDF}
\end{align}
where $\kappa_0$ is the saturation parameter at $x=0$.

\begin{figure}
\includegraphics{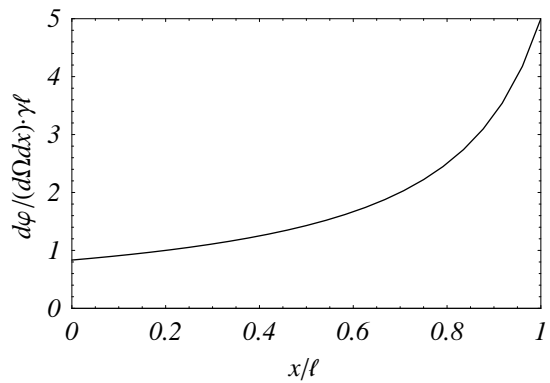}
\caption{Normalized contribution to small-field slope (in radians)
of a slice of the medium of width $dx$ at position $x$, given by
Eq. \ref{SlopeDxDF}. We have set $\kappa_0=1.8\alpha_0\ell$, the
value which is later seen to produce the greatest magnetometric
sensitivity.} \label{SlopeDxDFFig}
\end{figure}

\begin{figure}
\includegraphics{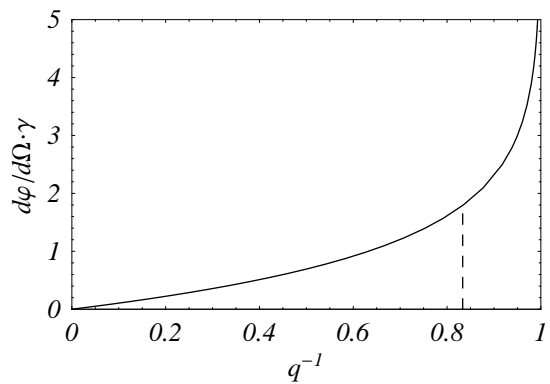}
\caption{Normalized slope (in radians) of NMOR as a function of
optical thickness of the medium
[$q^{-1}=(3/2)\alpha_0\ell/\kappa_0$]. The value of $q^{-1}$ that
produces the greatest magnetometric sensitivity is shown with a
dashed line.} \label{SlopeDFFig}
\end{figure}

The contribution to the small-field optical rotation of the
``slice'' of the medium at position $x$ is found by substituting
Eq.\ (\ref{kappaDF}) into Eq.\ (\ref{Slopedx}):
\begin{align}
    \frac{d\varphi}{d\Omega dx}(x)&\approx\frac{1}{\gamma}\frac{\alpha_0}{2\kappa_0/3-\alpha_0x}.
\label{SlopeDxDF}
\end{align}
This contribution is plotted as a function of position in Fig.
\ref{SlopeDxDFFig}. The plot illustrates that the part of the
medium near its end contributes significantly more to the overall
rotation than the part near its beginning. This is because light
power, and correspondingly, the light broadening of the resonance,
is lower at the end of the medium [the ratio of the rotation per
unit length is approximately
$\kappa(\ell)/\kappa_0\approx1-(3/2)\alpha_0 \ell/\kappa_0$].
Integrating over the length $\ell$ of the medium gives the total
slope:
\begin{align}
    \frac{d\varphi}{d\Omega}&=\int_0^\ell\frac{d\varphi}{d\Omega dx}(x)dx\nonumber\\
    &\approx\frac{1}{\gamma}\int_0^\ell\frac{\alpha_0}{2\kappa_0/3-\alpha_0x}dx\nonumber\\
    &=\frac{1}{\gamma}\ln\prn{\frac{\kappa_0}{\kappa_0-3\alpha_0\ell/2}}.
\label{SlopeDF}
\end{align}
The slope (\ref{SlopeDF}) is plotted as a function of
$\alpha_0\ell$ in Fig.\ \ref{SlopeDFFig}.

The photon shot-noise-limited magnetometric sensitivity $\delta B$
\endnote{In the general case, there may exist an additional source
of noise due to AC-Stark shift associated with off-resonant levels
\cite{Fle2000}; however, this source of noise is absent for an
isolated transition such as the one considered here.} is given in
terms of the number $N_\gamma$ of transmitted photons per unit
time, the slope of rotation with respect to $B$, and the
measurement time $t$:
\begin{align}
    (\delta B)^{-1}&=2\frac{d\varphi}{dB}\sqrt{N_\gamma t}\nonumber\\
    &=\frac{g\mu\omega}{\hbar c}\frac{d\varphi}{d\Omega}\sqrt{\frac{2At}{\pi}\gamma\kappa(\ell)}\nonumber\\
    &\approx\frac{g\mu\omega}{\hbar c}\sqrt{\frac{2At}{\pi}\frac{\kappa_0-\frac{3}{2}\alpha_0\ell}{\gamma}}\ln\prn{\frac{\kappa_0}{\kappa_0-3\alpha_0\ell/2}}\nonumber\\
    &=\frac{g\mu\omega}{\hbar c}\sqrt{\frac{3}{\pi}\frac{At\alpha_0\ell}{\gamma}}\sqrt{q-1}\ln\prn{\frac{q}{q-1}},
\end{align}
where $A$ is the cross-sectional area of the light beam and we
have made the change of variables $\kappa_0=3q\alpha_0\ell/2$. The
factor
\begin{align}
 \sqrt{q-1}\ln\prn{\frac{q}{q-1}}\nonumber
\end{align}
reaches a maximum of $\sim0.8$ at $q\approx1.2$ so we see that for
media of sufficient thickness, i.e.\ where $\alpha_0\ell\gg3$, and
for the optimum initial saturation parameter
$\kappa_0\approx1.8\alpha_0\ell$, we have from Eq.\
(\ref{kappaDF}) that $\kappa(x)\gg1$ for all $x$. For the optimum
light intensity, then,
\begin{align}
    (\delta B)^{-1} &\approx 0.8\frac{g\mu\omega}{\hbar
    c}\sqrt{\frac{At\alpha_0\ell}{\gamma}}\nonumber\\
    &\approx 1.1\frac{g\mu}{\hbar}\sqrt{\frac{A \ell n t}{\gamma}}.
\label{Sens_DF}
\end{align}
This result is consistent with the general observation
\cite{Bud2000Sens} that optimized NMOR provides a method for
measuring a spin-system at the standard quantum limit (SQL) given
by spin-projection noise \endnote{The use of so-called
spin-squeezed quantum states of light (see, e.g., Ref.\
\cite{Ula2001} and references therein) can, in principle, allow
one to overcome the SQL. We consider only non-squeezed states of
atoms and light here.}. The sensitivity is then expected to scale
as the square root of the product of the number of available atoms
and the spin-relaxation time (see, e.g., Ref.\ \cite{KhripLamor},
Sec.\ 3.1.1), which is indeed the result (\ref{Sens_DF}).

\section{The Doppler-broadened case}

Now we consider the case where the Doppler width $\Gamma$ is much
greater than the power-broadened line width. In this case, we need
to average over the over the atomic velocity distribution, which
is equivalent to averaging over the Doppler-free spectral
profiles. On resonance with the Doppler broadened transition, the
absorption coefficient is given by
\begin{align}
    \alpha_{DB}&\approx\frac{1}{\sqrt{\pi}\Gamma}\int_{-\infty}^{\infty}\alpha(\Delta)d\Delta\nonumber\\
    &\approx\frac{\alpha_0}{\sqrt{\pi}\Gamma}\int_{-\infty}^{\infty}\frac{d\Delta}{\prn{2\Delta/\gamma_0}^2+2\kappa/3}\nonumber\\
    & = \sqrt{\frac{3\pi}{8}}\frac{\gamma_0}{\Gamma}\frac{\alpha_0}{\sqrt{\kappa}}.
    \label{AlphaDB}
\end{align}
In this case the Doppler-broadened unsaturated absorbtion
coefficient is given in terms of the Doppler-free unsaturated
absorbtion coefficient by
\begin{align}
    \rightbar{\alpha_0}_{DB}=\frac{\sqrt{\pi}}{2}\frac{\gamma_0}{\Gamma}{\alpha_0}.
\end{align}
 Comparing Eqs.\ (\ref{AlphaDB}) and
(\ref{AlphaDF}), we see that we have reproduced a well-known
result (see, e.g., Ref.\ \cite{Dem96}, Sec.\ 7.2.1) that resonant
absorption falls as $1/\kappa$ for Doppler-free media and as
$1/\sqrt{\kappa}$ for Doppler-broadened media when $\kappa\gg1$.
The change in $\kappa$ per unit length is
\begin{align}
    \frac{d\kappa(x)}{dx}&=-\alpha_{DB}\kappa(x)\nonumber\\
    &\approx-\frac{\sqrt{6\pi}}{4}\frac{\gamma_0}{\Gamma}\alpha_0\sqrt{\kappa(x)},
\end{align}
and solving for $\kappa$ as a function of position,
\begin{align}
    \kappa(x)\approx\prn{\sqrt{\kappa_0}-\frac{\sqrt{6\pi}}{8}\frac{\gamma_0}{\Gamma}\alpha_0x}^2.
\label{KappaOfXDB}
\end{align}
Note that the behavior of the saturation parameter as a function
of distance is different in the Doppler-broadened case compared to
the Doppler-free case [cf.\ Eq.\ (\ref{kappaDF})] where the
saturation parameter falls approximately linearly with distance.

Taking the average of small-field rotation over the Doppler
distribution gives
\begin{align}
    \rightbar{\frac{d\varphi}{d\Omega
    dx}}_{DB}&\approx\frac{1}{\sqrt{\pi}\Gamma}\int_{-\infty}^{\infty}\frac{d\varphi}{d\Omega
    dx}\prn{\Delta}d\Delta\nonumber\\
    &\approx\frac{\alpha_0}{\sqrt{\pi}\gamma\Gamma}\int_{-\infty}^{\infty}\frac{1}{\prn{2\Delta/\gamma_0}^2+2\kappa/3}d\Delta\nonumber\\
    &\approx\frac{\sqrt{6\pi}}{4}\frac{\gamma_0}{\Gamma}\frac{\alpha_0}{\gamma\sqrt{\kappa}}.
    \label{SlopeDxDB}
\end{align}
We see that the rotation per unit length scales as
$1/\sqrt{\kappa}$, in contrast to the $1/\kappa$ scaling for the
Doppler-free case [similar to the situation with absorption, Eqs.\
(\ref{AlphaDF},\ref{AlphaDB})]. This is because the number of
atoms producing the effect is, in a sense, not fixed; with
increasing light power, a larger fraction of the Doppler
distribution is involved.

Substituting (\ref{KappaOfXDB}) into Eq.\ (\ref{SlopeDxDB}) to
find the contribution to the slope as a function of position gives
\begin{align}
    \rightbar{\frac{d\varphi}{d\Omega dx}(x)}_{DB}&\approx
    \frac{2}{\gamma}\frac{\alpha_0}{\frac{8}{\sqrt{6\pi}}\frac{\Gamma}{\gamma_0}\sqrt{\kappa_0}-\alpha_0x}.
    \label{SlopeDxOfXDB}
\end{align}
It is interesting to note that while the light power and the
rotation slope per unit length behave differently in the
Doppler-broadened case compared to the Doppler-free case, Eq.\
(\ref{SlopeDxOfXDB}) has the same functional form as for the
Doppler-free case [Eq.\ (\ref{SlopeDxDF})].

Integrating over the length of the medium, we obtain
\begin{align}
    \rightbar{\frac{d\varphi}{d\Omega}}_{DB}&\approx\int_0^\ell\rightbar{\frac{d\varphi}{d\Omega dx}(x)}_{DB}dx\nonumber\\
    &\approx\frac{2}{\gamma}\int_0^\ell\frac{\alpha_0}{\frac{8}{\sqrt{6\pi}}\frac{\Gamma}{\gamma}\sqrt{\kappa_0}-\alpha_0x}dx\nonumber\\
    &=\frac{2}{\gamma}\ln\prn{\frac{\sqrt{\kappa_0}}{\sqrt{\kappa_0}-\frac{\sqrt{6\pi}}{8}\frac{\gamma_0}{\Gamma}\alpha_0\ell}}.
\label{SlopeDB}
\end{align}
The behavior of (\ref{SlopeDB}) as a function of $\alpha_0\ell$ is
qualitatively similar to that of Eq.\ (\ref{SlopeDF}) shown in
Fig.\ \ref{SlopeDFFig}. However, the dependence on $\kappa_0$ is
different.

The magnetometric sensitivity is given by
\begin{align}
    &(\delta B)^{-1}_{DB}\nonumber\\
    &=\frac{g\mu\omega}{\hbar c}\rightbar{\frac{d\varphi}{d\Omega}}_{DB}\sqrt{\frac{2At}{\pi}\gamma\kappa(\ell)}\nonumber\\
    &\approx\frac{g\mu\omega}{\hbar c}\sqrt{\frac{8At}{\pi\gamma}}\prn{\sqrt{\kappa_0}-\frac{\sqrt{6\pi}}{8}\frac{\gamma_0}{\Gamma}\alpha_0\ell}\ln\prn{\frac{\sqrt{\kappa_0}}{\sqrt{\kappa_0}-\frac{\sqrt{6\pi}}{8}\frac{\gamma_0}{\Gamma}\alpha_0\ell}}\nonumber\\
    &=\frac{g\mu\omega}{\hbar c}\frac{\sqrt{3}}{2}\sqrt{\frac{At}{\gamma}}\frac{\gamma_0}{\Gamma}\alpha_0\ell\prn{p-1}\ln\prn{\frac{p}{p-1}},
\end{align}
with the change of variables
\begin{align}
    \kappa_0=\frac{3\pi}{64}\prn{p\frac{\gamma_0}{\Gamma}\alpha_0\ell}^2.
    \label{pDef}
\end{align}
The factor
\begin{align}
 \prn{p-1}\ln\prn{\frac{p}{p-1}}\nonumber
\end{align}
goes to unity as $p$ goes to infinity; it is approximately 0.9 at
$p=5$. (Further gain from increased power is minimal, and if the
power becomes too high, the approximation of the Doppler width
being much larger than the power-broadened width breaks down.)
Thus, in this case,
\begin{align}
    (\delta B)^{-1}_{DB}&\approx0.8\frac{g\mu\omega}{\hbar{}c}\sqrt{\frac{At}{\gamma}}\frac{\gamma_0}{\Gamma}\alpha_0\ell\nonumber\\
    &\approx0.3\frac{g\mu}{\hbar}\lambda\ell n \frac{\gamma_0}{\Gamma}\sqrt{\frac{At}{\gamma}}, \label{Sens_DB}
\end{align}
for sufficiently high $\kappa_0$.

This result, where sensitivity increases linearly with optical
thickness, holds for the case where the power-broadened width
$\sim\gamma_0\sqrt{\kappa(x)}$ is smaller than the Doppler width
for all $x$ within the sample, i.e.\
\begin{align}
1&\ll\kappa_0\ll\prn{\Gamma/\gamma_0}^2{\rm \ and}\nonumber\\
\Gamma/\gamma_0&\ll\alpha_0\ell\ll\prn{\Gamma/\gamma_0}^2,
\label{RestrictionsDB}
\end{align}
since $\alpha_0\ell$ is related to $\kappa_0$ by Eq.\
(\ref{pDef}). As power and optical thickness are increased beyond
this range, i.e.\
\begin{align}
\kappa_0&\gg\prn{\Gamma/\gamma_0}^2{\rm \ and}\nonumber\\
\alpha_0\ell&\gg\prn{\Gamma/\gamma_0}^2, \label{RestrictionsDF}
\end{align}
we obtain the Doppler-free case, where sensitivity increases as
the square root of the thickness [Eq.\ (\ref{Sens_DF})].

\section{The general case}

\begin{figure}[h!]
\includegraphics{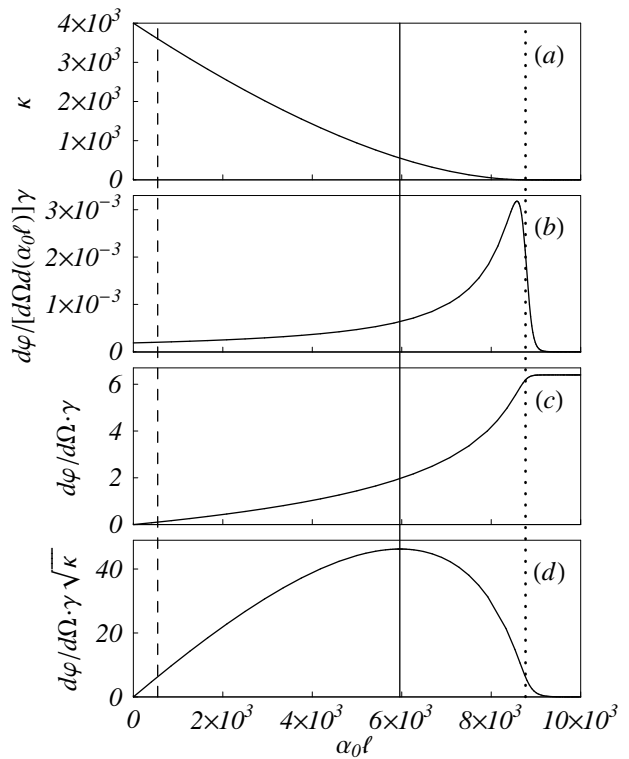}
\caption{Normalized (a) saturation parameter, (b) differential
small-field rotation, (c) total small-field rotation, and (d)
inverse magnetometric sensitivity as a function of optical depth
with initial saturation parameter $\kappa_0=4\times10^3$ and
$\Gamma/\gamma_0 = 60$. Plots (b), (c), and (d) are in units of
radians. The dashed line indicates the transition from the
Doppler-free to the Doppler-broadened regime
[$\kappa=\prn{\Gamma/\gamma_0}^2$], and the dotted line indicates
the point at which non-linear effects begin to turn off
($\kappa=1$). Linear optical rotation is neglected in this plot.
The solid line indicates the optical depth at which maximum
sensitivity is achieved.} \label{arbKappaFig}
\end{figure}

A numerical result can be obtained for the general case where the
restrictions (\ref{RestrictionsDB},\ref{RestrictionsDF}) on
$\kappa$ and $\alpha_0\ell$ are removed. In a typical experiment,
light power is $\sim1\ {\rm mW}$, the laser beam diameter is
$\sim0.1\ {\rm cm}$, $\lambda\approx800\ {\rm nm}$, and
$\Gamma/\gamma_0\approx60$. Thus the effective ground state
relaxation rate due to the transit of atoms through the laser beam
is $\gamma\approx2\pi\cdot50\ {\rm kHz}$ and the initial
saturation parameter is $\kappa_0\approx4\times10^3$. Here, as in
a typical experimental procedure, the optical depth is varied (by
changing atomic density) while the laser power is kept constant.
Normalized transmission, differential small-field rotation, total
small-field rotation, and magnetometric sensitivity are plotted in
Fig.\ \ref{arbKappaFig} as a function of optical depth.
For small
optical depth, $\kappa>\prn{\Gamma/\gamma_0}^2$ and the medium is
effectively Doppler-free. Transmission ($\propto\kappa$) falls
linearly until the transition to the Doppler-broadened case is
made (dashed line). Then transmission falls quadratically until
the linear regime is reached (dotted line), after which it falls
exponentially. Differential small-field rotation ($\propto
d\varphi/\sbrk{d\Omega d\prn{\alpha_0\ell}}\gamma$) initially
rises, as $\kappa$ falls and power broadening is reduced, until
non-linear effects begin to turn off. (Linear optical rotation is
neglected in this plot.) Since magnetometric sensitivity depends
both on total optical rotation and transmission, an intermediate
value for the optical depth produces the greatest sensitivity
(solid line). Multiplying the normalized inverse sensitivity
$d\varphi/d\Omega\cdot\gamma\sqrt{\kappa}$ by
\begin{align}
2\sqrt{2\pi}\frac{g\mu}{\hbar}\frac{\sqrt{A}}{\lambda\sqrt{\gamma}}\approx10^8\
\prn{{\rm G/\sqrt{Hz}}}^{-1} \label{geoFactor}
\end{align}
gives the absolute magnitude of sensitivity, $\sim2\times10^{-10}\
{\rm G/\sqrt{Hz}}$ (we assume $g=1$). Although this sensitivity is
not as high as could be achieved with low atomic density
paraffin-coated cells ($\sim3\times10^{-12}\ {\rm G/\sqrt{Hz}}$
\cite{Bud2000Sens}), it is, nevertheless, sufficiently high to be
of interest in practical applications \cite{Nov2002Mag}. In
particular, the power-broadening of the magnetic field dependence
of optical rotation at high light power provides an increased
dynamic range for magnetometry over the low-power case. There are,
however, techniques for shifting the narrow resonance obtained
with a paraffin-coated cell to higher magnetic fields, e.g.,
frequency modulation of the laser light \cite{Bud2002FM}.

\section{Large-field optical rotation}

\begin{figure}
\includegraphics{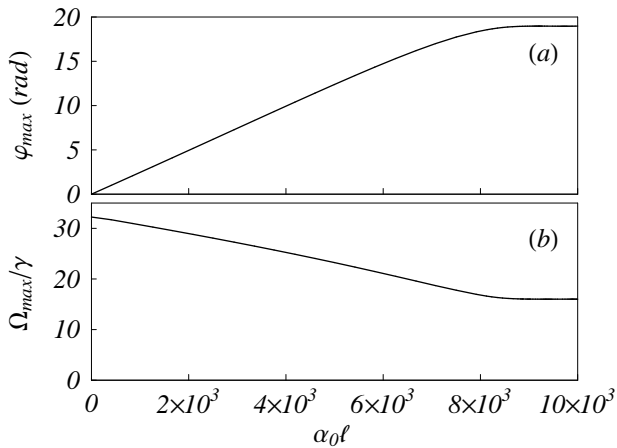}
\caption{(a) Maximum optical rotation and (b) normalized magnetic
field at which rotation is maximum as a function of optical depth.
Parameters are the same as those used for Fig.\
\ref{arbKappaFig}.} \label{MaxRotFig}
\end{figure}

The maximum optical rotation with respect to magnetic field can be
determined with a numerical calculation. Maximum rotation is
plotted as a function of optical depth in Fig.\ \ref{MaxRotFig},
for the same parameters as used for Fig.\ \ref{arbKappaFig}. The
rotation initially rises linearly, the same behavior seen in the
experiment \cite{NovLargePol2001} mentioned in the introduction.
For large optical depth, however, rotation in Fig.\
\ref{MaxRotFig} then begins to rise more quickly, and finally
saturates, whereas experimentally a slower increase and then a
decrease in rotation is seen. This is evidence for an additional
relaxation mechanism not accounted for in the present theory; in
Ref.\ \cite{NovLargePol2001} it is attributed to the effect of
radiation trapping.

\section{Conclusion}

In conclusion, we have analyzed magnetometric sensitivity of NMOR
measurements optimized with respect to light intensity in the case
of negligible Doppler broadening, and in the case of large Doppler
broadening. In the former case, we find that the sensitivity
improves as the square root of optical density, while in the
latter, it improves linearly. In the present discussion, we have
neglected the effect of velocity-changing collisions, which makes
this analysis not directly applicable to buffer-gas and
anti-relaxation-coated cells. However, since there is full mixing
between velocity components in these cells, one can expect that
the sensitivity should scale as square root of optical density (if
this quantity can be varied independently of the ground state
relaxation rate).

The authors are grateful to D. F. Kimball, I. Novikova, V. V.
Yashchuk, and M. Zolotorev for helpful discussions. This work has
been supported by the Office of Naval Research (grant
N00014-97-1-0214).

\bibliography{NMObibl}

\end{document}